\documentclass[manuscript]{aastex}
\usepackage{bm}
\usepackage{amsmath}

\begin{document}


\title{EVOLUTION OF SNOW LINE IN OPTICALLY THICK PROTOPLANETARY DISKS: EFFECTS OF WATER ICE OPACITY AND DUST GRAIN SIZE}


\author{Akinori Oka\altaffilmark{1}}
\email{akinorioka1@gmail.com}

\author{Taishi Nakamoto\altaffilmark{1}}
\email{nakamoto@geo.titech.ac.jp}

\and

\author{Shigeru Ida\altaffilmark{1}}
\email{ida@geo.titech.ac.jp}


\altaffiltext{1}{Department of Earth and Planetary Sciences, Tokyo Institute of Technology}


\begin{abstract}

Evolution of a snow line in an optically-thick protoplanetary disk is
investigated with numerical simulations.
The ice-condensing region in the disk is
obtained by calculating the temperature and the density with
the 1+1D approach.  The snow line migrates as the mass
accretion rate ($\dot{M}$) in the disk decreases with time.
Calculations are carried out from an early phase with high disk
accretion rates ($\dot{M} \sim 10^{-7} M_{\odot}$ yr$^{-1}$)
to a later phase with low disk accretion rates
($\dot{M} \sim 10^{-12} M_{\odot}$ yr$^{-1}$) 
using the same numerical method.
%
%
It is found that the snow line
moves inward for $\dot{M} \ga 10^{-10} M_{\odot}$ yr$^{-1}$,
while it gradually moves outward in the later evolution phase with
$\dot{M} \la 10^{-10} M_{\odot}$ yr$^{-1}$.
In addition to the silicate opacity, the ice opacity is taken into consideration.
In the inward migration phase,
the additional ice opacity increases the distance of the
snow line from the central star
by a factor of 1.3 for dust grains $\la 10$ $\mu$m in size and
1.6 for $\ga 100$ $\mu$m.
It is inevitable that the snow line
comes inside the Earth's orbit in the course of the disk evolution,
if the alpha viscosity parameter
is in a range 0.001-0.1, the dust-to-gas mass ratio is higher than a tenth of the
solar abundance value, and the dust grains are smaller than 1 mm.
The formation of water-devoid planetesimals in the terrestrial
planet region seems to be difficult throughout the disk evolution,
which imposes a new challenge to planet formation theory.

\end{abstract}





%
%
\section{INTRODUCTION}

In a protoplanetary disk, a snow line which is defined as the inner boundary of an ice-condensing region is present. 
The snow line is considered to play an important role on planetary formation,
since the solid mass density outside the snow line is high due to the condensation of water ice
(water ice abundance is as large as silicate and iron in a protoplanetary disk with the solar composition [Pollack et al. (1994)]). 
The snow line may also be related to the origin of the water distribution in our solar system. 
Water abundance in the current solar system objects shows a clear variation; bodies in the inner part contain less, while bodies in the outer part have more. 
The terrestrial planets are significantly devoid of water. 
For example, Earth contains only 0.023wt\% water in its ocean, and
Venus is considered to have contained a comparable amount of water with Earth (Lewis 2004). 
On the other hand, the outer planets and objects outside the asteroid belt contain a large amount of water. 
In the intermediate region, i.e., the asteroid belt, the radial distribution of asteroids also shows compositional zoning. 
From these facts, one may consider that the snow line might have been located in the asteroid belt when the planetesimal formation occurred, since planetesimals which were formed outside the snow line would necessarily take in a large amount of icy materials. 

The location of the snow line, which is used in many studies on planet formation, is the heliocentric distance where the temperature of the disk becomes about 170 K. 
In the solar nebula, the snow line location is estimated to be 2.7 AU assuming that the solar nebula is optically thin  (Hayashi 1981). 
Before the planet formation takes place, however, protoplanetary disks are likely to be optically thick, since a large amount of fine dust particles are present.
Thus, temperatures in optically thick disks should be obtained to determine the location of the snow line. 

So far, the snow line location in an optically thick disk has been theoretically obtained by 1D (radial) or 1+1D (radial and vertical) disk structure calculations (e.g., Cassen 1994; Stepinski 1998). 
The latest studies on the snow line location with a detailed disk structure calculation were done by Davis (2005) and Garaud \& Lin (2007). 
They considered the stellar radiation flux and the viscous dissipation of gas as the main heating sources in a disk, and obtained disk temperature by solving the detailed radiative energy transfer. 
They revealed that the snow line migrates inward as the disk evolves and the mass accretion rate decreases,
because the viscous dissipation of gas, which is the main heating source in the disk, reduces as the disk evolves.
Davis (2005) showed that the snow line reaches about 0.6 AU, which is the minimum radius in
his calculations.
Similarly, Garaud \& Lin (2007) found that in the later phase,
the snow line migrates outwardly since the stellar radiation penetrates deeper into the disk interior
as the disk becomes optically thinner and rises the temperature.


One important point of their results is that the snow line comes inside the Earth orbit; the minimum heliocentric distance of the snow line is about 0.6 AU (Davis 2005; Garaud \& Lin 2007). 
If sufficiently large bodies like planetesimals were formed when the snow line was located at such a small heliocentric distance, Earth would have been formed with icy planetesimals. 
Then, Earth should presently contain a comparable amount of water with silicate and iron. 
This conflicts with the current water content in Earth. 


In order to make clear the important inconsistency in planet formation, 
we investigate the detailed thermal evolution of protoplanetary disks
using precise radiative transfer calculations
taking into account the frequency dependence on opacity, 
the scattering process of dust particles,
and the ice opacity that previous studies did not consider.
According to Inoue, Oka, \& Nakamoto (2009) and Dullemond {\it et al}. (2002), disk midplane temperature varies by a factor of about two at most due to the first two effects. 
Ice opacity may change the optical property of a disk or disk temperature structure significantly,
since ice is as abundant as silicate in a protoplanetary disk of solar composition.
%
And, the effect of the dust grain size,
which changes the snow line location as well,
has not received enough consideration in previous studies.
These effects are examined carefully in this work to see the location of the snow line.

In this study, the snow line location is numerically simulated with the 1+1D disk model taking into account
the water ice opacity, the variety of dust grain size, and the scattering process in the radiative transfer.
Then, the theoretically obtained snow line location is discussed if it is consistent with the water distribution in the current solar system. 


%
%
\section{MODEL}

\subsection{Disk structure}
A geometrically thin axisymmetric disk revolving around a central star is considered.
A plane symmetry with respect to the midplane is assumed.
Cylindrical coordinates $(R, \phi, Z)$ are adopted.
The origin and the $Z=0$ plane are put to the central star and the disk midplane.

The hydrodynamical and thermal structures of the disk are modeled almost the same way as Dullemond {\it et al}. (2002) and Inoue, Oka, \& Nakamoto (2009) except for the disk's gas surface density $\Sigma (R)$. 
The radiative cooling and two heating sources, which are the irradiation by the central star and the viscous dissipation, are considered.
The temperature is determined with the radiative equilibrium condition. 
To obtain the whole disk structure, the 1+1D approach is adopted.
The direct irradiation from the central star at the disk inner edge is ignored, and the disk structure is assumed to change gradually in the radial direction. 
This approach is justified if the snow line location is sufficiently far from the inner edge. 
The 2D structure of the inner disk region (the puffed-up inner rim and the adjacent shadowed region) seems to be important only in the innermost part of the disk around a T Tauri star judging from results for Herbig Ae/Be stars by Dullemond (2002). 
Though the shadowed region extends to slightly large heliocentric distances, it does not matter for the scope of this study because it only lowers the disk temperature and does not shift the snow line location outward. 
Details are given in Appendix. 

The radial distribution of the gas surface density is modeled assuming that the disk is in the steady accretion state, i.e., the mass accretion rate $\dot{M}$ is constant along the radial direction $R$. 
The gas surface density in a region where $R \gg R_{*} $ (the stellar radius)  is given as (e.g., Pringle 1981),
\begin{equation}
	\Sigma(R) = \frac{\dot{M}}{3\pi \nu_{t}(R)},
	\label{eq:steady_accretion}
\end{equation}
where $\nu_{t}(R)$ is the viscosity in the disk. 
The assumption of the steady accretion is justified in the region around the snow line 
because the snow line is usually located within $10$AU and the viscous diffusion timescale there is sufficiently smaller than the entire disk evolution timescale. 

The viscosity $\nu_{t}$ is modeled with the $\alpha$-prescription (Shakura \& Sunyaev 1973) as
$ \nu_{t} = \alpha c_{s}^{2}/\Omega_{\mathrm{K}}$,
where $\alpha$ is a non-dimensional parameter, $c_{s}=\sqrt{kT / (\bar{\mu} m_{u})}$ is the isothermal sound velocity of gas, $k$ is the Boltzmann constant, $T$ is the gas temperature, $\bar{\mu}$ is the mean molecular weight, $m_{u}$ is the atomic mass unit, 
$\Omega_{\mathrm{K}} = \sqrt{GM_{*} / R^{3}}$ is the Kepler angular velocity, $G$ is the gravitational constant, and $M_{*}$ is the stellar mass, respectively. 
The value of $\alpha$ ranges from $0.001$ to $0.1$ according to ideal MHD simulations (Hawley {\it et al}. 1995, 1996). 
The value of $\alpha = 0.01$ is adopted as a fiducial one throughout the disk in this study except for \S 3.3. 
In the evaluation of $\nu_{t}$, the sound velocity $c_{s}$ at the midplane ($Z=0$) is used. 
As for the mean molecular weight, $\bar{\mu} = 2.3$ is used. 

The frequency dependent radiative transfer is solved numerically. 
Absorption and scattering by dust particles are taken into consideration as the opacity source in the disk. 
Isotropic scattering is assumed, when the dust grain size is sufficiently smaller than the radiation wavelength. 
When the dust grain size is larger than the wavelength, the scattering coefficient is set to be zero, because a forward scattering dominates in that case.

The steady state disk structure is solved self-consistently as a function of the mass accretion rate. 
Though density, viscosity, and temperature are related to each other, they are numerically obtained consistently by iterative calculations. 

\subsection{Dust abundance and opacity}

Only dust particles suspended in the disk are considered to be the opacity source because 
the gas opacity is negligibly small. 
Two kinds of dust particles, silicate and pure water ice particles, are assumed to be present separately. 
Effects of icy mantles on silicate cores will be discussed in \S 5.2. 
It is assumed that dust particles are uniform-sized spheres and
mass ratios of the silicate and ice particles to the gas are constant throughout the disk. 
A radial transport, a vertical mixing, and the sedimentation of dust particles are neglected. 
Also, a complete thermal coupling between dust particles and gas is assumed,
which is valid in the disk considered in this study (Kamp \& Dullemond 2004).

\begin{figure}
\epsscale{1.0}
\plotone{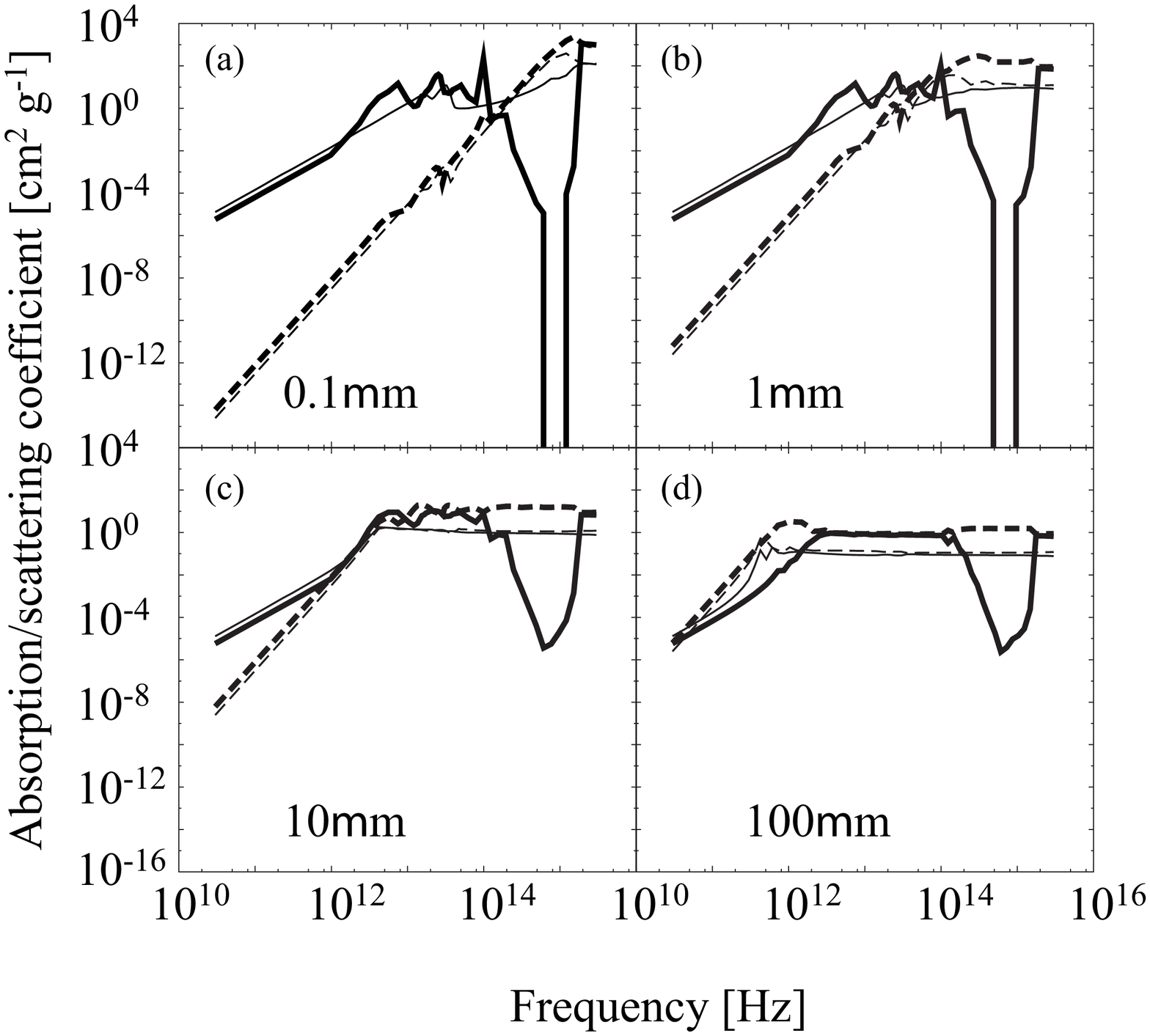}
\caption{}
\label{opacity_table}
\end{figure}

Absorption and scattering coefficients of dust particles,
and mass ratios to the gas are taken from Miyake \& Nakagawa (1993);
the silicate and the (maximum) ice mass ratios to the gas are
$\zeta_{\mathrm{sil}} = 0.0043$ and $\zeta_{\mathrm{ice}} = 0.0094$, respectively,
which are consistent with the solar elemental abundance given by Anders \& Grevesse (1989). 
With this water ice abundance and the mean molecular weight of 2.3, $\mathrm{H_{2}O}$ mole fraction in the disk gas becomes $1.2 \times 10^{-3}$. 
Figure \ref{opacity_table} shows the absorption and scattering coefficients of a unit mass disk medium. 

The ice abundance in the disk gas is determined with the thermal equilibrium condition under which the partial pressure of water vapor is limited by the saturated vapor pressure (Bauer {\it et al}. 1997): 
\begin{equation}
	P_{\mathrm{sat}}(T) = \exp \left( -6070\mathrm{K} / T + 30.86 \right)\ \mathrm{dyn \, cm^{-2}}. 
	\label{eq:partial_pressure}
\end{equation}
The partial pressure of water vapor $P_{\mathrm{H_{2}O}} $ is approximately given by the product of the water vapor mole fraction $X_{\mathrm{H_{2}O}}$ and the total gas pressure: 
\begin{equation}
	P_{\mathrm{H_{2}O}} = X_{\mathrm{H_{2}O}} P. 
\end{equation}
The water vapor mole fraction $X_{\mathrm{H_{2}O}}$ is limited by $1.2 \times 10^{-3}$, so excess water molecules condense as water ice. 
Thus, $X_{\mathrm{H_{2}O}}$ is given by 
\begin{equation}
	X_{\mathrm{H_{2}O}} = 
	\begin{cases}
		1.2 \times 10^{-3}
		& (1.2 \times 10^{-3} P \le P_{\mathrm{sat}}(T)), \\
		P_{\mathrm{sat}}(T)/ P
		&  (1.2 \times 10^{-3} P > P_{\mathrm{sat}}(T)).
	\end{cases}
\end{equation}
Then, the mass ratio of ice to the total water molecules in the disk gas, $x_{\mathrm{ice}}$, is given by 
\begin{equation}
	x_{\mathrm{ice}}  = 1 - X_{\mathrm{H_{2}O}} / \left(1.2 \times 10^{-3} \right) .
\end{equation}

Using $x_{\mathrm{ice}}$, the absorption and scattering coefficients of the disk medium, $\kappa_{\mathrm{abs}}$ and $\kappa_{\mathrm{sca}}$, are given as 
\begin{eqnarray}
	\kappa_{\mathrm{abs}} &=& \kappa_{\mathrm{sil, abs}} + x_{\mathrm{ice}} \kappa_{\mathrm{ice, abs}}, 
	\label{kappa_1} \\
	\kappa_{\mathrm{sca}} &=& \kappa_{\mathrm{sil, sca}} + x_{\mathrm{ice}} \kappa_{\mathrm{ice, sca}}, 
	\label{kappa_2}
\end{eqnarray}
where $\kappa_{\mathrm{sil, abs}}$, $\kappa_{\mathrm{sil, sca}}$, $\kappa_{\mathrm{ice, abs}}$, and $\kappa_{\mathrm{ice, sca}}$ 
represent the absorption and scattering coefficients of disk medium attributed to silicate and ice dust particles, respectively. 
Eqs. (\ref{kappa_1}) and (\ref{kappa_2}) are held for each frequency of the radiative transfer. 
The disk temperature depends on $x_{\mathrm{ice}}$, hence the temperature and $x_{\mathrm{ice}}$ are solved consistently by iterative calculations. 

%
%
\section{RESULTS}

\subsection{Effects of water ice opacity}


Numerical results shown in this subsection are obtained by calculations with following input parameters:
$T_{\mathrm{eff}}=3000$ $\mathrm{K}$, $R_{*}=2$ $R_{\odot}$, $M_{*}=0.5$ $M_{\odot}$,
and the dust grain size is $0.1$ $\mathrm{\mu m}$. 

\subsubsection{Disk structure}

Figures \ref{temperature_profile8_0} and \ref{temperature_profile10_0} show vertical temperature profiles at various radii with $\dot{M}$ $=$ $10^{-8}M_{\odot}\mathrm{yr}^{-1}$ and $10^{-10}M_{\odot}\mathrm{yr}^{-1}$, respectively, as typical cases for high and low mass accretion rates. 
In Fig. \ref{temperature_profile8_0}, the temperature in the disk with the ice opacity is higher than without it. 
The dominant heating source in this case is the viscous dissipation, and the blanket effect by dust particles is enhanced due to the ice in upper layers. 
On the other hand, in Fig. \ref{temperature_profile10_0},
the midplane temperatures are not different each other considerably though ice condenses. 
In this case, the dominant heating source is the incident stellar radiation, so the midplane temperature does not depend on the dust opacity so much. 
The temperature in the disk is lowered due to the ice condensation.
The low temperature causes a slightly low altitude of the scale height and $H_{s}$ as seen in Fig. \ref{temperature_profile10_0}
($H_{s}$ is the height from the midplane where the incident stellar radiation energy flux reduces by a factor of $e$.) 

The disk temperature is altered only a little by considering the dust scattering,
because a single scattering albedo for 0.1 $\mu$m-sized dust particles is small.
Though it is small, the midplane temperature is increased slightly when the viscous heating is dominant
(Fig. \ref{temperature_profile8_0});
the scattering by dust particles enhances the optical depth in the disk and increases the blanket effect.
On the contrary, the midplane temperature is slightly lowered when the stellar irradiation is the dominant heating source (Fig. \ref{temperature_profile10_0}), since a large fraction of incident radiation is reflected toward space (Dullemond et al. 2003; Inoue, Oka, \& Nakamoto 2009).

\subsubsection{Evolution of ice-condensing region}

Figure \ref{2D_snowline} shows the ice-condensing region in the $(R, Z)$ sectional plane with various mass accretion rates\footnote{
The flaring index $\xi$ is adopted to be zero at small heliocentric distances ($R\lesssim 4\mathrm{AU}$)
for $\dot{M}\gtrsim 4\times 10^{-8}M_{\odot}\mathrm{yr}^{-1}$, because the grazing angle in an inner region becomes negative due to a high temperature caused by the viscous heating, and the grazing angle recipe becomes invalid. 
Though this operation is artificial, the disk structure around the midplane is not affected significantly by the operation thanks to the fact that the dominant heating source around the midplane is the viscous dissipation. 
This operation is also done if the grazing angle becomes negative in the other calculations in this study. 
}. 
A decrease of mass accretion rate can be regarded as the time evolution of the disk. 

When $\dot{M}$ is high ($\ge 10^{-9}M_{\odot}\mathrm{yr}^{-1}$), 
the condensation front has a two-branched structure as found by Cassen (1994) and Davis (2005). 
The condensation front in the upper layer extends closer to the central star than the snow line (hereafter, the snow line is defined as the condensation front at the midplane) irrespective of the presence of ice. 
The lower branch of the condensation front is formed by a high temperature caused by the viscous dissipation, 
while the upper branch is formed by a high temperature caused by the direct stellar radiation and a low partial pressure of water vapor. 
The snow line shifts outward by a factor of 1.3 due to the ice opacity independent of the mass accretion rate. 

On the other hand, when $\dot{M}$ is low ($\le 10^{-10}M_{\odot}\mathrm{yr}^{-1}$), 
the condensation front is a smooth curve and the snow line location is not shifted significantly by the ice opacity. 
This can be attributed to two reasons: when the disk is mainly heated by the stellar irradiation,
(1) the midplane temperature does not depend on the opacity of the disk medium so much, and 
(2) icy dust particles do not condense in the upper layer above the snow line due to the temperature increasing with height. 

\subsubsection{Evolution of snow line}

Figure \ref{snowline_evolution} shows the snow line locations obtained with and without ice opacity as a function of mass accretion rate. 
In both cases, the snow line migrates inward when $\dot{M}$ is high ($\gtrsim 10^{-10}M_{\odot}\mathrm{yr}^{-1}$), while it migrates outward when $\dot{M}$ is low ($\lesssim 10^{-10}M_{\odot}\mathrm{yr}^{-1}$). 
An inward migration is caused by the decrease of the midplane temperature due to the reduction of the viscous heating and the decrease of the blanket effect. 
The inward migration ceases when the stellar irradiation starts to dominate the heating around the midplane. 
Then, an outward migration is started by the decrease of the disk's optical thickness which results in a higher temperature due to a deeper penetration of the stellar radiation into the disk interior. 
The snow line in this study migrates outward more slowly than the results by Garaud \& Lin (2007)
(see \S 3.3). 

When $\dot{M}$ is high ($\gtrsim 10^{-10}M_{\odot}\mathrm{yr}^{-1}$),
the ratio of the snow line distance with ice opacity ($R_{\rm SL, sil+ice}$)
and that without ice opacity ($R_{\rm SL, sil}$), $f_{\rm SL} = R_{\rm SL, sil+ice} / R_{\rm SL, sil}$,
is about 1.3 and almost constant.
This constant ratio of the snow line shift factor $f_{\rm SL}$ is originated from assumptions that the disk is the steady accretion disk and 
the mass fraction of ice dust particles to silicate dust particles is uniform throughout the disk
(see \S 4.1).
On the other hand, when $\dot{M}$ is low ($\lesssim 10^{-10}M_{\odot}\mathrm{yr}^{-1}$)
the snow line location is not substantially affected by the ice opacity (i.e., $f_{\rm SL} \simeq 1$). 
This small difference in the snow line locations is caused by
a small difference in the disk structure around the snow line. 

\subsection{Dependence of snow line location on dust grain size}

Stellar parameters in this subsection are set as $L_{*} = 1L_{\odot}$, $M_{*} = 1M_{\odot}$, and $T_{\mathrm{eff}} = 4000\mathrm{K}$ for comparison with the solar system. 
Figure \ref{solar_parameter_snowline} shows the heliocentric distance of snow line as a function of mass accretion rate. 
Several curves correspond to different dust grain sizes assumed in each calculation.
Calculations are stopped when the optical depth in the region around the snow line for the stellar radiation becomes less than unity. 
When the size of grains is $\geq 100$ $\mathrm{\mu m}$, the scattering coefficient is set to be zero. 

During the inward migration of the snow line, the snow line location is almost independent of the dust grain size if the grain size is smaller than 10 $\mu$m. 
The wavelength of radiation from the disk interior at the condensation temperature ($T \simeq 170\mathrm{K}$) is longer than 10 $\mu$m,
so the opacity of the dust particle is in the Rayleigh regime, dependent only on the total mass of the dust particles, and dominated by the absorption.
On the other hand, when the size is larger than 10 $\mathrm{\mu m}$, i.e., in the geometrical optics regime, the opacity depends on the total cross section of the dust particles and decreases with the dust grain size, and hence the temperature enhancement by the viscous heating becomes inefficient,
although both absorption and scattering contribute in this case. 



When the dust grain size is larger than 10 $\mu$m,
the evolutional track of the snow line levels off at the minimum value,
because
the opacity for large grains is almost independent of the wavelength emitted from the disk. 
This implies that the snow line necessarily passes the heliocentric distance of 1 AU,
unless the dust grain size exceeds a certain large value. 
This point will be considered in \S 4.4. 

The shift ratio of the snow line $f_{\rm SL}$ is qualitatively the same as in Figure \ref{snowline_evolution}. 
The snow line shifts to a larger heliocentric distance in its inward migration phase, whereas it does not shift substantially in its outward migration and leveling-off phases. 

The shift ratio of the snow line $f_{\rm SL}$ depends on the dust grain size:
$f_{\rm SL} \simeq 1.3$ for the dust grain size $\lesssim$ 10 $\mu$m,
and $f_{\rm SL} \simeq 1.6$ for the grain size $\gtrsim$ 100 $\mu$m. 
When the dust grain size is small, the total dust opacity increases in proportion to the total mass of the dust,
while the total dust opacity increases in proportion to the total cross section of dust particles when the dust grain
size is large.
This different dependence of the opacity on the total amount of dust causes the different dependence of the shift ratio.

The critical mass accretion rate, that is
the mass accretion rate with which the snow line starts to migrate outward,
also depends on the dust grain size.
The snow line starts to migrate outward when the disk region around the snow line becomes transparent.
When the dust grain size is sufficiently larger than the wavelength of the stellar radiation ($\sim 1 \mathrm{\mu m}$),
the total opacity of dust is inversely proportional to the grain size.
Thus, the critical mass accretion rate is proportional to the dust grain size.
Indeed, when the dust grain size is 10 $\mu$m, 100 $\mu$m, and 1 mm,
the snow line starts to migrate outward at about $10^{-11}M_{\odot}\mathrm{yr^{-1}}$,
$10^{-10}M_{\odot}\mathrm{yr^{-1}}$,
and $10^{-9}M_{\odot}\mathrm{yr^{-1}}$, respectively (Fig. \ref{solar_parameter_snowline}).

\subsection{Comparison with Garaud \& Lin (2007)}

The snow line evolution obtained by our numerical simulations is compared with that by the semi-analytical model of Garaud \& Lin (2007).
The stellar parameters are $L_{*} = 1L_{\odot}$, $M_{*} = 1M_{\odot}$, and $R_{*} = 1R_{\odot}$,
and the viscosity parameter is $\alpha = 0.001$.
Other settings are almost the same as Garaud \& Lin (2007).

Garaud \& Lin (2007) used the Planck mean opacity,
thus a new frequency dependent opacity model of the mixture of the disk gas and dust particles is developed,
so that its Planck mean agrees with that of Garaud \& Lin (2007). 
The Planck mean opacity is proportional to the temperature, so the absorption opacity is modeled as a linear function of the frequency: $
	\kappa_{\nu}^{\mathrm{abs}} = C \nu
$
where $C$ is a constant determined by 
\begin{equation}
	\frac{ \int_{0}^{\infty} \kappa_{\nu}^{\mathrm{abs}} B_{\nu}(T_{\mathrm{eff}} )d\nu }{ \int_{0}^{\infty} B_{\nu}(T_{\mathrm{eff}} )d\nu } = \kappa_{V}, 
\end{equation} 
where $\kappa_{V}$ is the Planck mean opacity with the stellar effective temperature, and $\kappa_{\rm V}$ is set to be $2$ $\mathrm{ cm^{2}\ g^{-1}}$ according to Garaud \& Lin (2007).

Calculated snow line location as a function of  $\dot{M}$ is shown in Figure \ref{Garaud_Lin_snowline}, as well as the result by Garaud \& Lin (2007).
Our result agrees well with that by Garaud \& Lin (2007) for higher $\dot{M}$ ($ > 10^{-9}M_{\odot}\mathrm{yr}^{-1}$), though
our result does not show an abrupt outward shift of the snow line at $\dot{M} \sim 10^{-9.5}M_{\odot}\mathrm{yr}^{-1}$.
Garaud \& Lin (2007) adopted a semi-analytical method to solve the temperature in the optically-thin region. 
Although semi-analytical methods are much easier to use than numerical models, they may not be accurate enough for the evaluation of the disk midplane temperature in optically-thin regions.
(They are accurate enough and useful in optically-thick regions.)
The discontinuous evolution found by Garaud \& Lin (2007) may come from the simplified treatment of the radiative transfer. 
Hereafter, the discontinuity is ignored and overall features of the snow line evolution is discussed.

In order to obtain the midplane temperature, the estimation of the height of the super heated layer $H_{s}$ is a key factor, because its radial dependence determines the grazing angle of the disk for irradiation.
When the midplane temperature at a region is increased, the surface density there is lowered according to the steady accretion assumption (eq. (\ref{eq:steady_accretion})).
Then, the grazing angle decreases, and the midplane temperature tends to lower and the surface density tends to increase: a negative feed back takes place. 
Therefore, in our calculations, the disk does not become optically so thin and the snow line migrates outward slowly than the result by Garaud \& Lin (2007), in which $H_{s}$ is assumed to be proportional to the disk pressure scale height. 

The minimum heliocentric distance of the snow line obtained by our calculations is slightly smaller than that of Garaud \& Lin (2007). 
This may be attributed to the two layer model (Chiang \& Goldreich 1997) they adopted. 
Inoue, Oka, \& Nakamoto (2009) showed that the three layer approach is necessary to obtain an accurate midplane temperature. 

Figures \ref{temperature_Garaud_Lin} and \ref{surface_density_Garaud_Lin} show the midplane temperature and the surface density profiles by our calculation and by Garaud \& Lin (2007). 
These figures clearly show that
the midplane temperatures by both models agree well in the inner optically-thick region,
whereas our result is lower than that by Garaud \& Lin (2007) in the outer optically-thin region.
Similarly, the surface density distributions in both models agree in the inner region,
whereas there is a difference in the outer region. 

We conclude that if we are concerned with detailed evolution of the snow line in the optically-thin phases in which planet formation may proceed, we need to numerically calculate the temperature.
(If we are concerned with optically-thick phases or qualitative features of the snow line evolution in entire phases, semi-analytical calculations are sufficient.)

%
%
\section{Semi-analytical estimates}


\subsection{Dependence on water ice opacity}


In the early phase in which the snow line migrates inward,
the main heating source at the midplane around the snow line is the viscous dissipation (Fig. \ref{temperature_profile8_0}). 
So, the temperature is determined by the viscous heating and the radiative cooling. 
The diffusive radiative energy transfer gives the midplane temperature, $T_{c}$, as (e.g., Nakamoto \& Nakagawa 1994),
\begin{equation}
	T_{c} = \left[ \left( \frac{3}{2}+\frac{9\tau_{c}}{16} \right) \left( \frac{GM_{*}\dot{M}}{4\pi \sigma R^{3}} \right) \right]^{\frac{1}{4}},
	\label{eq:midplane_temperature}
\end{equation}
where $\tau_{c}$ is the optical depth for the midplane and $\sigma$ is the Stefan-Boltzmann constant. 
Substituting $\tau_{c} =\kappa \Sigma/2$ ($\kappa$ is the Rosseland mean opacity of the disk medium) and noticing that $\tau_{c}$ is much larger than unity, we obtain,
\begin{eqnarray}
	T_{c} \simeq \left[ \left(\frac{9\kappa \Sigma}{32} \right) \left( \frac{GM_{*}\dot{M}}{4\pi \sigma R^{3}} \right) \right]^{\frac{1}{4}} 
	\propto \left( \frac{\kappa\Sigma}{R^{3}} \right)^{\frac{1}{4}}. 
\end{eqnarray}
The surface density of the steady accretion disk is given by $\Sigma=\dot{M}/(3\pi\alpha c_{s}^{2}/\Omega)\propto 1/(T_{c}R^{3/2})$.
So, we have:
\begin{equation}
	T_{c}^{5} \propto \kappa R^{-\frac{9}{2}}. 
	\label{eq:T_c}
\end{equation}

In the annulus located at the snow line, both silicate and ice particles contribute to the opacity $\kappa$ 
(see Figure \ref{2D_snowline}). 
This Rosseland mean opacity is denoted as $\kappa_{\mathrm{sil+ice}}$, whereas 
the Rosseland mean opacity only due to silicate particles is expressed as $\kappa_{\mathrm{sil}}$. 
Since the ice condensation temperature at the midplane is almost independent of the heliocentric distance, 
the right hand side of eq. (\ref{eq:T_c}) can be regarded as constant, and then a relation between the snow line locations with and without the water ice opacity, $R_{\mathrm{SL, sil+ice}}$ and $R_{\mathrm{SL, sil}}$, is obtained as 
\begin{equation}
	\kappa_{\mathrm{sil+ice}} R_{\mathrm{SL, sil+ice}}^{-\frac{9}{2}} = \kappa_{\mathrm{sil}} R_{\mathrm{SL, sil}}^{-\frac{9}{2}}.
\end{equation}
The shift factor of the snow line location due to the additional ice is then written as, 
\begin{equation}
	f_{\rm SL} =
	\frac{R_{\mathrm{SL, sil+ice}}}{R_{\mathrm{SL, sil}}} = \left( \frac{\kappa_{\mathrm{sil+ice}}}{\kappa_{\mathrm{sil}}} \right)^{\frac{2}{9}}.
	\label{radial_shift_snow_line}
\end{equation}

Adopting the dust opacity model by Miyake \& Nakagawa (1993), we have $\kappa_{\mathrm{sil+ice}}/\kappa_{\mathrm{sil}} \simeq 3$ for the thermal radiation with 170K (the midplane temperature at the snow line) when the dust grain size is 0.1 $\mathrm{\mu m}$. 
Then, $f_{\rm SL}$ is evaluated as $f_{\rm SL} = 3^{\frac{2}{9}} = 1.28$,
which is consistent with our numerical results. 
Equation (\ref{radial_shift_snow_line}) implies that the snow line location is shifted outward in accordance with the water abundance around the snow line. 

\subsection{Dependence on viscosity parameter}

The snow line location is affected by $\alpha$ considerably,
because $\alpha$ relates to the disk's optical thickness and energy generation rate by the viscous dissipation.
In the early phase of the disk evolution, the midplane temperature in the inner disk region is obtained in the same
way as \S4.1. 
Then, we have $
	T_{c}^{5} \propto \alpha^{-1} R^{-9/2}. 
$
And we find
the dependence of the snow line location $R_{\mathrm{snow}}$ on $\alpha$ as 
\begin{equation}
	R_{\mathrm{snow}} \propto  \alpha^{-\frac{2}{9}}. 
\end{equation}
Taking into account the range of $\alpha$ ($0.001 \lesssim \alpha \lesssim 0.1$), 
the snow line location shifts inward or outward by a 67 \% at most from the results with $\alpha = 0.01$ in the inward migration phase. 
Note that the minimum heliocentric distance of the snow line hardly changes, because it is determined not by the viscous heating but by the stellar radiation heating. 

In the later phase of the disk evolution, a difference of $\alpha$ changes the correlation between $R_{\rm snow}$ and $\dot{M}$. 
In the outward migration phase, $R_{\rm snow}$ is a function of $\Sigma$, and $\dot{M}$ is proportional to the product of $\nu_{\rm t}$ and $\Sigma$. 
The viscosity $\nu_{\rm t}$ is proportional to $\alpha$, so $\dot{M}$ corresponding to each $R_{\rm snow}$ varies in proportion to $\alpha$. 
Thus, $R_{\rm snow}$ in the later phase is shifted in proportion to $\alpha$ (Figure \ref{solar_parameter_snowline}). 
This means that $\dot{M}$ with which the disk becomes optically thin for the stellar radiation also decreases in proportion to $\alpha$. 

\subsection{Dependence on dust-to-gas mass ratio}

The dust-to-gas mass ratio, $\zeta_{d}$, changes the opacity of disk medium. 
Hence it changes the disk temperature in the same way as ice opacity. 
In the inward migration phase,
assuming that the opacity of disk medium varies in proportion to $\zeta_{d}$,
a proportional relation between $R_{\mathrm{snow}}$ and $\zeta_{d}$ is obtained
in the same way as \S 4.1 and \S 4.2:
\begin{equation}
	R_{\mathrm{snow}} \propto \zeta_{d}^{\frac{2}{9}}. 
\end{equation}

In the outward migration phase,  the correlation between $R_{\rm snow}$  and $\dot{M}$
depends on $\zeta_{d}$. 
In this phase, $R_{\rm snow}$ is a function of solid mass surface density ($\zeta_{d} \Sigma$),
and $\dot{M}$ is proportional to $\Sigma$ which is the product of solid mass surface density and an inverse of $\zeta_{d}$. 
Thus, $\dot{M}$ corresponding to each $R_{\rm snow}$ varies in inverse proportion to $\zeta_{d}$.
Hence, $R_{\rm snow}$ is shifted inversely proportional to $\zeta_{d}$ (Figure \ref{solar_parameter_snowline}). 
The mass accretion rate with which the disk becomes optically thin for the stellar radiation increases in inverse proportion to $\zeta_{d}$. 

\subsection{Possibility of snow line coming inside Earth's orbit}

According to our results shown in \S3, the snow line comes inside the current Earth's orbit in the course
of the disk evolution. However, if planetesimals in the terrestrial planet region were icy, it is not consistent with
current solar system bodies.
Here, discussing the snow line evolution track on $(\dot{M}, R_{\mathrm{snow}})$-plane in Fig. \ref{solar_parameter_snowline},
we examine to what extent that conclusion is certain.


From Fig. \ref{solar_parameter_snowline}, we can see that if $(\alpha / \zeta_{d})$ is increased by more than a factor of 100 and if the dust grain size is 1 mm, the snow line would not come inside Earth's orbit.
However, there are some caveats. 
Noticing that a range of $\alpha$ is from 0.001 to 0.1 and the fiducial value used in our calculations is $\alpha$ = 0.01, we realize that $\zeta_{d}$ should be decreased at least by a factor of 10 from the value by Miyake \& Nakagawa (1993), to increase the value of $(\alpha /\zeta_{d})$ by more than a factor of 100. 
When $\alpha$ $< 0.1$, $\zeta_{d}$ should be much small in accordance with the smallness of $\alpha$.  
On the other hand, as the dust grain size becomes small, $(\alpha / \zeta_{d})$ should be large (or $\zeta_{\mathrm{sil}}$ should be small). 
Thus, it seems that the snow line necessarily comes inside the Earth orbit, if the dust grain size is smaller than 1 mm, $\zeta_{d}$ is higher than a tenth of that of the solar abundance, and $\alpha$ ranges from 0.001 to 0.1.

Another possibility is an increase of ice abundance; it can prevent the snow line from coming inside the Earth's orbit. 
The increase of ice abundance lowers $\dot{M}$ with which the snow line migration levels off at the minimum distance or starts to move outwardly. 
For example, if the water ice mass ratio to the disk gas $\zeta_{\mathrm{ice}}$ is increased by a factor of $10^{4}$, it is expected that the snow line would not come inside the Earth's orbit even in the case in which $\alpha = 0.01$, the dust grain size is 1 mm, and $\zeta_{\mathrm{sil}} = 0.0043$ (the solar abundance). 
When the dominant dust grain size or $(\alpha / \zeta_{d})$ is small, a more increase of the ice abundance is needed. 

%
%
\section{DISCUSSION}

\subsection{Comparison with the current solar system}

In our solar system, the water distribution shows a drastic change at the asteroid belt;
this may be a clue of the snow line. 
Terrestrial planets are thought to form from planetesimals
which have formed inside the snow line, otherwise a large amount of water is inevitably accumulated
into planets. 
Then, the timing of planetesimal formation may be restricted by the snow line evolution. 

First, we consider the possibility that planetesimals form after the snow line leaves outwardly the terrestrial planet region. 
When we look at the solid mass, the planetesimal formation with a large sized dust grains is favorable, because the gas surface density with which the snow line turns its direction of motion and moves outwardly increases as the dust grain size increases (Figure \ref{solar_parameter_snowline}). 
As the most favorable case, we see the numerical result of the 1 mm-sized dust case. 
Our numerical simulation shows that the surface density of solid material at the heliocentric distance of 1 AU is $1.1\times 10^{-2}$ $\mathrm{g\ cm^{-2}}$ when the outwardly moving snow line reaches 1 AU. 
This is much smaller than that of the minimum mass solar nebula model (about $10$ $\mathrm{ g\ cm^{-2}}$). 
To match the minimum mass solar nebula model, or to match the current terrestrial planets' mass, the size of the dust particles would be necessarily larger than 1 m. 
This would be unrealistic,
partly because a complete depletion of fine dust particles smaller than 1 m would be difficult,
and partly because 1 m-sized dust particles are eliminated efficiently by accretion toward the central star due to the gas drag. 
Thus, the planetesimal formation during the snow line's outward migration would be unrealistic. 

Next, let us consider the possibility of planetesimal formation during the snow line's inward migration. 
If planetesimal formation completes before the inwardly-migrating snow line reaches the terrestrial planet region,
planetesimals devoid of water can be formed.
As for the solid mass, planetesimal formation in this phase is favorable.
However, as the grain size grows, the snow line migrates fast inwardly and eventually comes inside the Earth's orbit;
in Figure \ref{solar_parameter_snowline}, at a fixed $\dot{M}$, we can see that $R_{\rm snow}$ is small if the grain size is large.
Hence, the dust grain growth and planetesimal formation should be completed in a short period of time compared to the timescale with which the snow line moves inward. 
If not, a large amount of ice would necessarily accumulate deeply into planetesimals. 
And we are not sure if the planetesimal formation completed in such a short period of time.

In conclusion, the snow line location during the disk evolution does not seem to match the current solar system. 
The inconsistency may originate from the following assumptions adopted in this study: 
(1) the continuous boundary condition at the inner disk edge, (2) uniform dust grain size and uniform dust-to-gas mass ratio throughout the disk, and (3) neglecting the evolution of solid material and water distributions. 

Here, we consider possible scenarios by relaxing above assumptions.

(1) Continuous disk boundary:
the presence of the inner disk edge, where the disk is heated by a direct stellar radiation and becomes hot, is ignored by this assumption. 
If the dust grain grows at the inner edge and the inner edge moves outward after the planetesimal formation, the dust grain growth would always proceed in a hot environment. Then the resulting planetesimals would be devoid of water.

(2) Uniform dust grain size and dust-to-gas mass ratio:
In a real disk, the dust grain size and the dust-to-gas mass ratio would be non-uniform. 
Dust particles grow mainly by collisions. And the growth tends to proceed from the inner disk region, so the dust grain size in the inner disk is likely to be larger than that in the outer disk.
When the dust grain is large enough, the height of the disk surface is lowered.
Consequently, a disk region, which is located just outside the developed dust grain region and the dust grain growth is about to start, can receive more stellar radiation flux. 
Then, the dust grain growth may proceed with a hot temperature and water deficit planetesimals may form. 

(3) Solid material and water distribution evolution:
According to Ciesla \& Cuzzi (2006) and Garaud (2007), a large amount of water vapor and fine dust particles are transported from the outer part to the inner part of the disk during the disk evolution. 
If fine dust particles or water vapor are supplied to the terrestrial planet region, and the snow line location is kept farther from the terrestrial planet region until the planetesimal formation is completed, the dust grain growth may proceed in a water ice-free environment, and consequently water-devoid planetesimals may be formed. 
Whether or not this mechanism works may depend on the initial condition of the disk. 

Yet another possible scenario may be an elimination of water from icy planetesimals after the snow line leaves outwardly the terrestrial planet region.
Future work on these points is needed.

\subsection{Effects of ice mantle on silicate dust particles}

In this study, it is assumed that ice condenses as pure ice particles and no ice mantle forms on silicate particles. 
It would be more realistic, however, that $\mathrm{H_{2}O}$ molecules condense as ice mantles on silicate particles. 
Here, we discuss how the ice mantle formation affects the snow line location.
We consider its effects only in the inward migration phase, since the ice opacity does not change the snow line location in the outward migration phase. 

To evaluate effects of the ice mantle, we consider to what extent the dust opacity changes due to the ice mantle. 
Properties of the dust opacity change as the ratio between the dust grain size and the wavelength of the thermal radiation ($\lambda \sim 10$ - $20\mathrm{\mu m}$) changes.  
When the dust grain size is sufficiently smaller than 10 - 20 $\mathrm{\mu m}$ (the Rayleigh regime), 
the total dust opacity is hardly affected by the ice mantle. 
On the contrary, when the dust grain size is sufficiently larger than 10 - 20 $\mathrm{\mu m}$ (the geometrical optics regime), the ice mantle changes the total dust opacity. 

When the ice condenses, the opacity of dust particles generally increases.
However, the opacity enhancement factor $g_{\rm ice}$ can be different between cases in which the ice condenses as independent ice particles and in which the ice condenses as ice mantles on silicate cores, even though the amount of condensed ice is the same.
In the geometrical optics regime, $g_{\rm ice}$ for the mantle case becomes smaller than that for the independent case, since the cross section of ice mantles is smaller than that of the independent particles.
According to Miyake \& Nakagawa (1993), the material densities of silicate and ice are 3.3 and 0.92 $\mathrm{g\ cm^{-3}}$, respectively, and the mass fractions of silicate and water ice in the disk medium are 0.0043 and 0.0094, respectively. 
This means that ice has about 6 times larger volume than silicate in a unit mass disk medium. 
As a result, if only pure ice particles are formed, 
the total cross section of the dust particles is increased by a factor of 7.9,
while if ice mantles are formed on silicate cores, 
the total cross section of dust particles is increased only by a factor of $3.7$. 
Thus, the ice mantle formation reduces the opacity enhancement factor by the ice condensation. 
Consequently, the snow line's shift ratio by the ice condensation ($f_{\rm SL}$) would be reduced from $1.54$ to $1.33$ due to the ice mantle formation (eq.(\ref{radial_shift_snow_line})). 

Finally, we consider the nonuniformity of the dust grain size. 
When independent ice particles are formed, the opacity enhancement depends on the size distribution of ice particles. 
If all the ice particles are sufficiently small compared to the wavelength of the thermal radiation (the Rayleigh regime), the opacity enhancement depends only on the total mass of ice and becomes the largest.
When the dominant size of ice particles is larger than 10 - 20 $\mathrm{\mu m}$,
and as the dominant size increases, the opacity enhancement decreases. 
On the other hand, if ice mantles are formed on silicate particles, the opacity enhancement would be limited in a certain range. 
A real situation should be somewhere between the Rayleigh regime and the geometrical optics regime, so the enhancement of the dust opacity by the ice condensation would be around 30\%. 

In summary, it is likely that the enhancement of the dust opacity would be limited to a range around 30\%, when ice mantles are formed around silicate dust particles. 
If pure ice particles are formed, the size distribution of ice particles is needed to evaluate the enhancement accurately. 

%
%
\section{CONCLUSIONS}

The evolution of the snow line in an optically-thick disk is numerically simulated and examined if it is consistent with the water distribution in the solar system. 
The evolution is examined from the early phase in which the mass accretion rate in the disk is large and the
optical depth of the disk is considerably high to the later phase in which the accretion rate is small and the optical
depth is low with a single numerical scheme.

The snow line migrates as the mass accretion rate in the disk decreases (which is considered to be the evolution).
Generally, in the early phase (high mass accretion phase), it moves inwardly because of the reduce of the
viscous heating, while in the later phase (low mass accretion phase), it moves outwardly due to the stellar radiation.
If the dust grain size is large ($\gtrsim 10\mathrm{\mu m}$), the snow line stays at its minimum heliocentric distance for a wide range of the mass accretion rate. 

When the opacity by the condensed ice is taken into consideration,
we have found that the snow line location is shifted outwardly
for the disk in the inward migration phase ($\dot{M} \gtrsim 10^{-10}M_{\odot}\mathrm{yr^{-1}}$). 
This is due to the additional blanket effect by the condensed ice particles in the upper layer of the disk. 
The shift ratio of the snow line ($f_{\rm SL} $ $=$ $ R_{\rm SL,ice+sil}/R_{\rm SL,sil}$) varies with the dust grain size: $f_{\rm SL}= 1.3$ for the grain $\lesssim 10\mathrm{\mu m}$ and $f_{\rm SL} = 1.6$
 for the grain $\gtrsim 100\mathrm{\mu m}$. 
Our semi-analytical estimation has shown that $f_{\rm SL}$ increases with the water abundance in the disk gas around the snow line. 
However, the snow line shift due to the ice opacity is small compared to the total migration length during the disk evolution, and is limited in the early phase of the disk evolution. 

The additional ice opacity does not change the snow line location of the disk in the outward migration phase. 
A vertically increasing temperature profile under the irradiation by the central star
prevents water molecules from condensing in the upper layer of the disk.

We have also found that the snow line comes inside the Earth's orbit as long as the dust-to-gas mass ratio is higher than about a tenth of the solar abundance, the viscosity parameter $\alpha$ is $0.001\lesssim \alpha \lesssim 0.1$, and the dust grain size is smaller than 1 mm. 
Then, if one thinks that terrestrial planets should be formed from water-devoid planetesimals, the dust grain growth should occur either before the snow line comes inside the Earth's orbit or after the snow line passes outward the Earth's orbit. 
In the latter case, the formation of water-devoid planetesimals is impossible because of the deficit of solid mass
(\S5.2). 
In the former case, the dust grain growth should be completed within 1 yr because the snow line migrates inwardly with this timescale. 

The inconsistency of the snow line evolution obtained in this study with the solar system would originate from assumptions of this study: (1) the continuous boundary condition at the inner disk edge, (2) uniform dust grain size and uniform dust-to-gas mass ratio throughout the disk, and (3) neglecting the evolution of solid material and water distributions. 
Future work on these points is needed.

\acknowledgments

We are grateful to Masahiro Ikoma and Hidekazu Tanaka for their fruitful discussions and useful comments. 
We also thank an anonymous reviewer for helpful comments.

\appendix

\section{Appendix}

The density and the temperature of the disk are obtained using the 1+1 D approach. 
In this approach, one divides the disk into many annuli and solves the vertical structure of each annuli, and then adds them into the entire (radial and vertical) structure of the disk. 
Although we solve the dynamical and thermal equilibria separately, we calculate them iteratively to obtain the consistent solution. 

\subsection{Hydrostatic equilibrium}

The hydrostatic equilibrium along the vertical direction is expressed as,
\begin{equation}
	\frac{\partial P(R,Z)}{\partial Z} = -\rho(R,Z)\frac{GM_{*}}{R^{3}}Z,
\end{equation}
where $P$ is the gas pressure and $\rho$ is the mass density of the gas. 
The equation of state of the gas is given as,
$P = {\rho kT \over \bar{\mu} m_{u}}= c_{s}^{2} \rho$, which $\bar{\mu}$ is the mean molecular weight
$m_u$ is the atomic mass unit, and $c_s$ is the sound velocity.
Integrating the density over $Z$, the surface density $\Sigma$ is given as
$	\Sigma = \int_{-\infty}^{\infty}\rho \, dZ$.

\subsection{Radiative transfer}


Two heating sources are taken into consideration: the viscous dissipation and the irradiation of the central star. 
Balancing the heating sources and the radiative cooling, the equilibrium temperature is obtained:
\begin{equation}
	q_{\mathrm{vis}} + q_{\mathrm{irr}} + \int_{0}^{\infty}\rho\kappa_{\nu}(1-\varpi_{\nu}) d\nu \oint I_{\nu}(\bm{\Omega})d\Omega 
	= 4\pi \int_{0}^{\infty}\rho\kappa_{\nu}(1-\varpi_{\nu})B_{\nu}(T)d\nu,
	\label{eq:radiative_equilibrium}
\end{equation}
where $q_{\mathrm{vis}} $ is the heating rate by the viscous dissipation, $q_{\mathrm{irr}}$ is the heating rate by the irradiation of the central star, $\kappa_{\nu}$ is the extinction coefficient, $\varpi_{\nu}$ is the single scattering albedo, $I_{\nu}(\bm{\Omega})$ is the diffuse radiation field in the disk (intensity without including the direct radiation from the central star), 
and $B_{\nu}(T)$ is the Planck function. 
Quantities with subscript $\nu$ are a function of the frequency $\nu$.
The integral $\oint d\Omega$ on the left hand side of eq. (\ref{eq:radiative_equilibrium}) represents the integral over the entire solid angle. 
The viscous heating rate $q_{\mathrm{vis}}$ is given by (e.g., D'Alessio et al. 1998),
$ 
	q_{\mathrm{vis}} = \frac{9}{4}\rho\nu_{t}\Omega_{\mathrm{K}}^{2}.
$ 

The diffuse component of the intensity $I_{\mu,\nu}(\bm{\Omega})$ and $q_{\mathrm{irr}}$ are obtained by solving the radiative transfer. 
Assuming a plane-parallel structure along the $Z$-direction, the radiative transfer equation is written as
\begin{equation}
	\mu \frac{dI_{\mu,\nu}(Z)}{dZ} = \rho \kappa_{\nu} (S_{\nu}(Z) - I_{\mu,\nu}(Z)),
	\label{eq:radiative_transfer}
\end{equation}
where $\mu$ is the cosine of the propagation direction of radiation
and $S_{\nu}$ is the source function.
The source function $S_{\nu}$ is given by 
\begin{equation}
	S_{\nu}(Z) = (1-\varpi_{\nu}(Z)) B_{\nu}(T) + \varpi_{\nu}(Z) \frac{1}{2}\int_{-1}^{1}I_{\mu,\nu}d\mu 
	+ \varpi_{\nu}(Z)\frac{F_{\mathrm{irr},\nu}(Z)}{4\pi},
	\label{eq:source_function}
\end{equation}
where $F_{\mathrm{irr},\nu}(Z)$ is the radiation energy flux from the central star. 
Therein, the isotropic scattering is assumed. 
The energy flux $F_{\mathrm{irr},\nu}(Z)$ is evaluated using the so-called grazing angle recipe (e.g., Chiang \& Goldreich 1997; Dullemond {\it et al}. 2002): 
\begin{equation}
	F_{\mathrm{irr},\nu}(Z) = \frac{L_{\nu}}{4\pi R^{2}} \exp\left(-\frac{\tau_{\nu}(R,Z)}{\beta} \right),
\end{equation}
where $\beta$, $L_{\nu}$, and $\tau_{\nu}(R,Z)$ are the cosine of the penetrating angle of the radiation from the central star into 
the disk surface, the luminosity of the central star, 
and the optical depth between a point $(R,Z)$ and the infinity along the vertical direction $(R, \infty)$, respectively. 
The optical depth $\tau_{\nu}$ is defined as
$ 
	\tau_{\nu}(R,Z) = \int_{Z}^{\infty}\kappa_{\nu}\rho(R,Z) dZ.
$ 
The central star is assumed to emit the black body radiation with the effective temperature $T_{\mathrm{eff}}$,
so the luminosity is given by $L_{\nu} = 4\pi^{2} R_{*}^{2} B_{\nu}(T_{\mathrm{eff}})$. 
And $\beta$ is evaluated by (e.g., Kusaka et al. 1970; Chiang \& Goldreich 1997; Dullemond et al. 2002),
\begin{equation}
	\beta(R) = 0.4\frac{R_{*}}{R}+ \xi(R)\frac{H_{s}}{R}, 
\end{equation}
where $H_{s}$ is the height from the midplane where the radiation from the central star loses its energy by $1-(1/e)$ 
and $\xi$ is the so-called flaring index defined as  
$ 
	\xi = \frac{d \ln (H_{s}/R)}{d \ln R}.
$ 
Then, the heating rate by the irradiation of the central star in eq. (\ref{eq:radiative_equilibrium}) is given by
\begin{equation}
	q_{\mathrm{irr}} = \int_{0}^{\infty}\rho(Z)\kappa_{\nu}(1-\varpi_{\nu})F_{\mathrm{irr},\nu}(Z)d\nu.
\end{equation}
The boundary condition for eq. (\ref{eq:radiative_transfer}) is 
\begin{eqnarray}
	&&I_{\mu, \nu}(+\infty) = 0 \ (\mu <0), \\
	&&I_{\mu, \nu}(0) = I_{-\mu, \nu}(0) \ (\mu \ge 0).
\end{eqnarray}
This condition represents the mirror boundary at the $Z=0$ plane and no incident diffuse radiation at the disk surface. 

Vertical profiles of the diffuse radiation $I_{\mu,\nu}(\bm{\Omega})$ and the temperature $T$ are obtained by solving eqs. (\ref{eq:radiative_equilibrium}), (\ref{eq:radiative_transfer}), 
and (\ref{eq:source_function}) iteratively. 
A straight forward way to solve these equations converges very slowly, so the variable Eddington factor (VEF) method, described in Dullemond {\it et al}. (2002) and Inoue, Oka, \& Nakamoto (2009), is employed.
In this study, the VEF method is modified to deal with the heating by the viscous dissipation.

\newpage
\section{Figure Captions}

Fig. 1.--- Absorption and scattering coefficients of the disk gas containing silicate and icy dust particles adopted in our calculation. 
These coefficients are taken from Miyake \& Nakagawa (1993). 
Note that all the water molecules are condensed in the icy dust particles. 
In the panels (a)-(d), the dust grain size is set to be 0.1 $\mathrm{\mu m}$, 1 $\mathrm{\mu m}$, 10 $\mathrm{\mu m}$, and 100 $\mathrm{\mu m}$, respectively. 
The thick and thin curves represent the coefficients by icy and silicate dust particles, respectively, and 
the solid and dashed curves represent the absorption and scattering coefficients, respectively. 

Fig. 2.--- Vertical temperature profiles at various heliocentric distances. The mass accretion rate is fixed to be $10^{-8}M_{\odot}\mathrm{yr}^{-1}$. 
Other parameters are $M_{*} = 0.5$ $M_{\odot}$, $R_{*} = 2$ $R_{\odot}$, $T_{\mathrm{eff}} = 3000$ K, $\alpha = 0.01$, and the dust grain size is 0.1 $\mathrm{\mu m}$. 
Panels (a)-(d) show the profiles at the heliocentric distances of 0.5, 1, 2, and 4 AU, respectively. 
The horizontal axis represents the height from the midplane normalized by the heliocentric distance. 
In each panel, the upper part shows the temperature profile, 
and the lower part shows the vertical profile of the ice-condensing ratio. 
The solid and dotted curves show the results with and without scattering by dust particles, respectively. 
The black and gray curves show the results with and without the ice opacity. 
The solid circle and the solid square on each curve show the height of $H_{s}$ and the pressure scale height, respectively. 

\begin{figure}
\epsscale{1.0}
\plotone{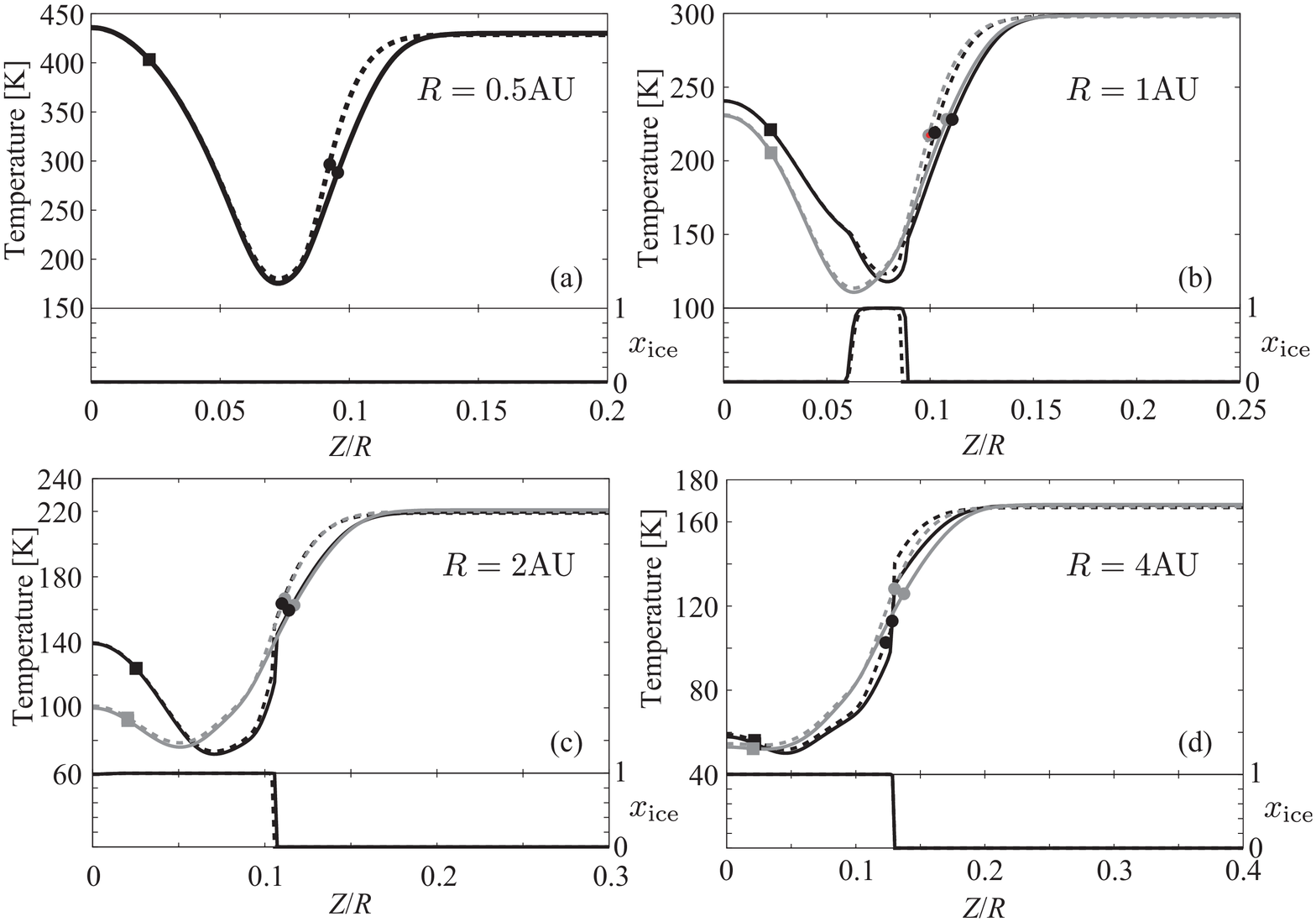}
\caption{}
\label{temperature_profile8_0}
\end{figure}

\begin{figure}
\epsscale{1.0}
\plotone{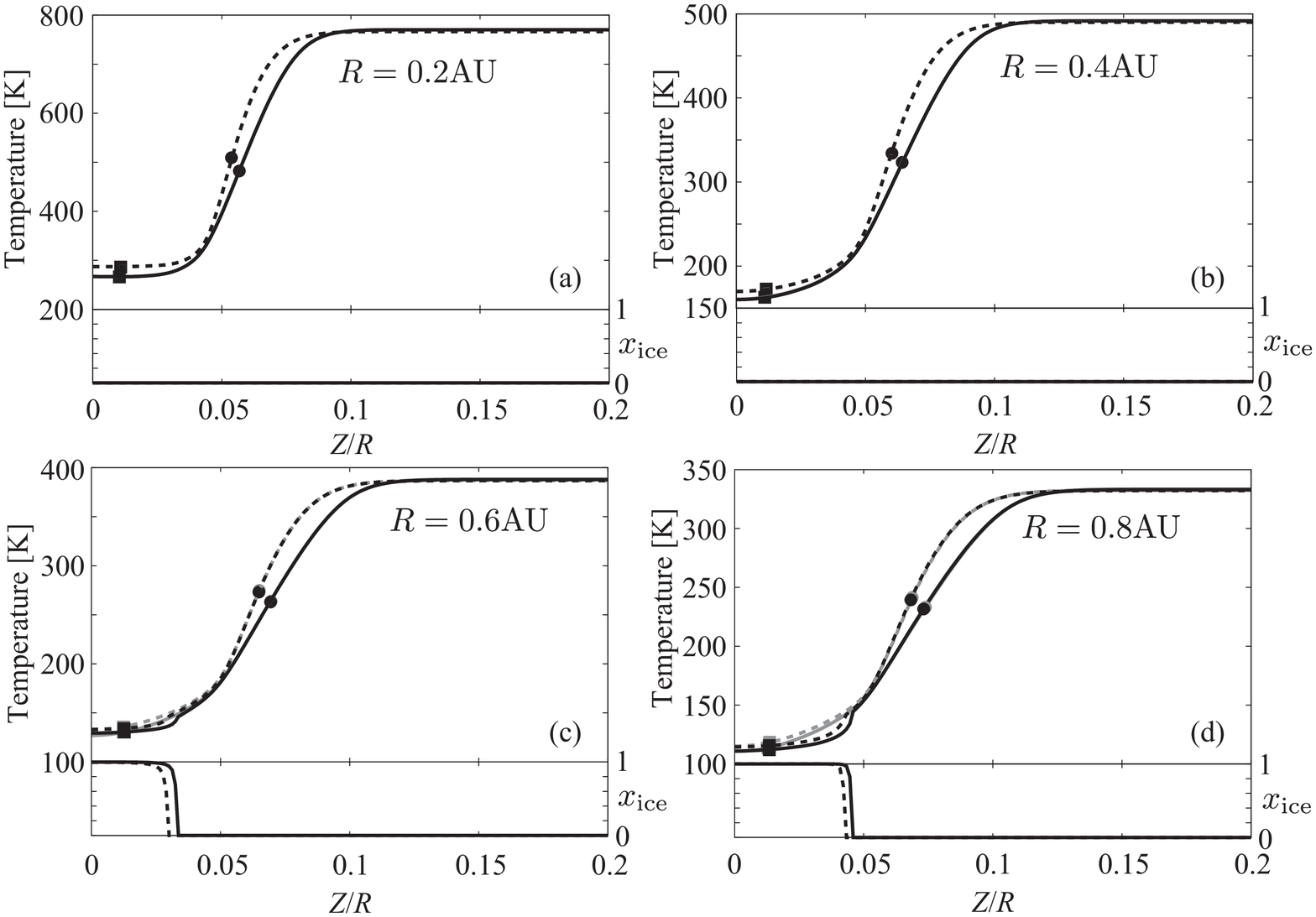}
\caption{}
\label{temperature_profile10_0}
\end{figure}

\begin{figure}
\epsscale{1.0}
\plotone{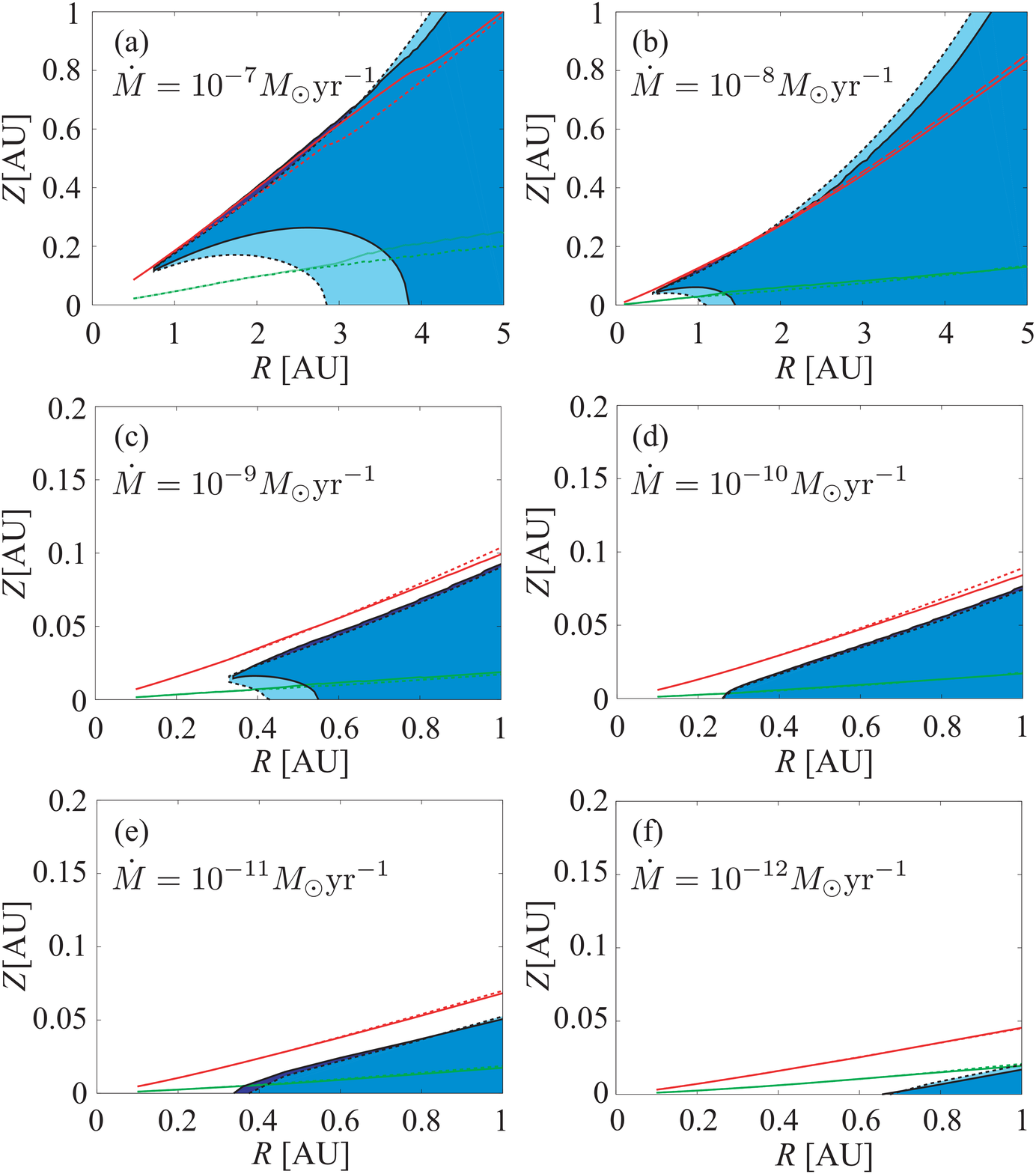}
\caption{}
\label{2D_snowline}
\end{figure}

\begin{figure}
\epsscale{1.0}
\plotone{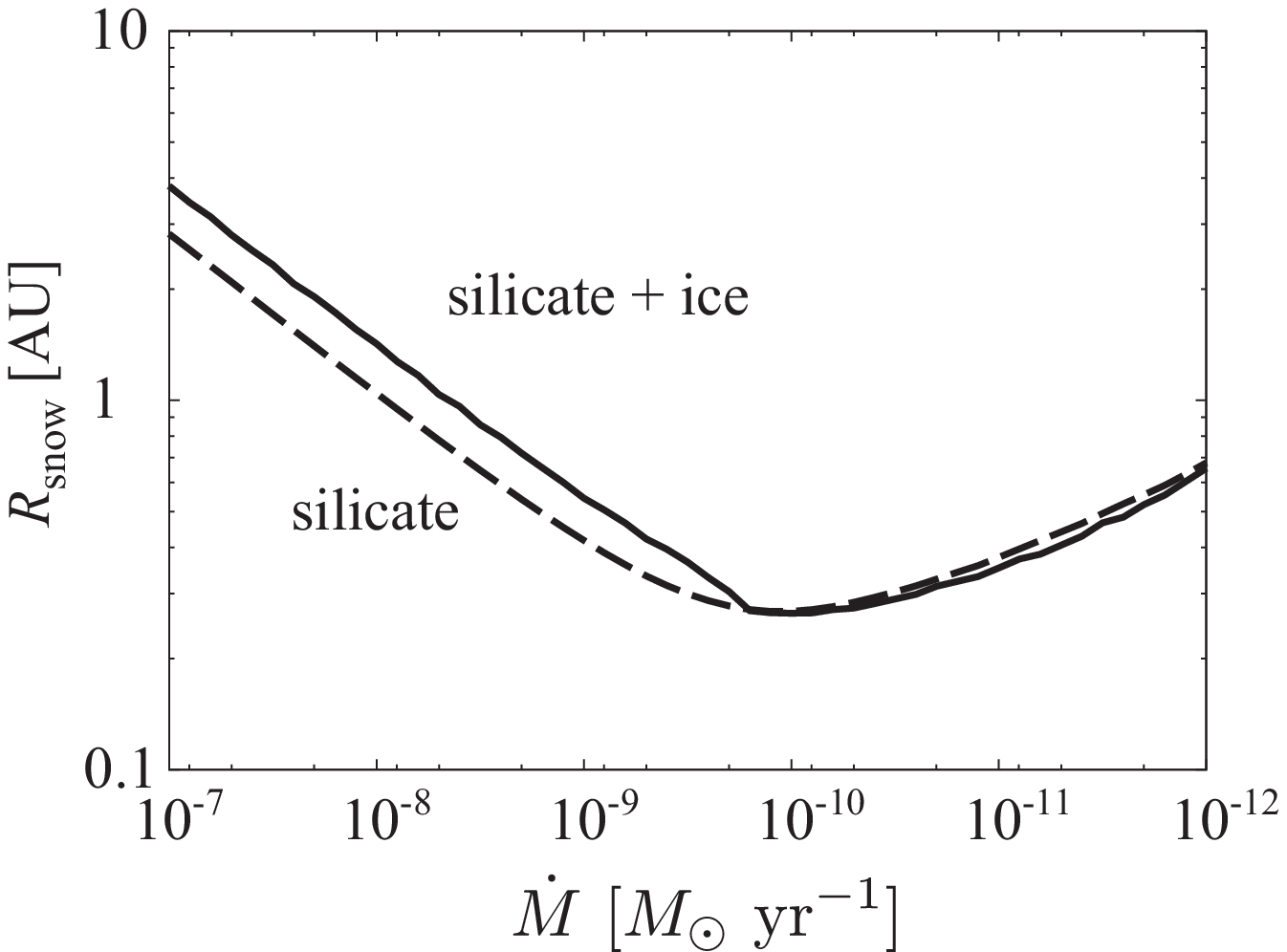}
\caption{}
\label{snowline_evolution}
\end{figure}

\begin{figure}
\epsscale{1.0}
\plotone{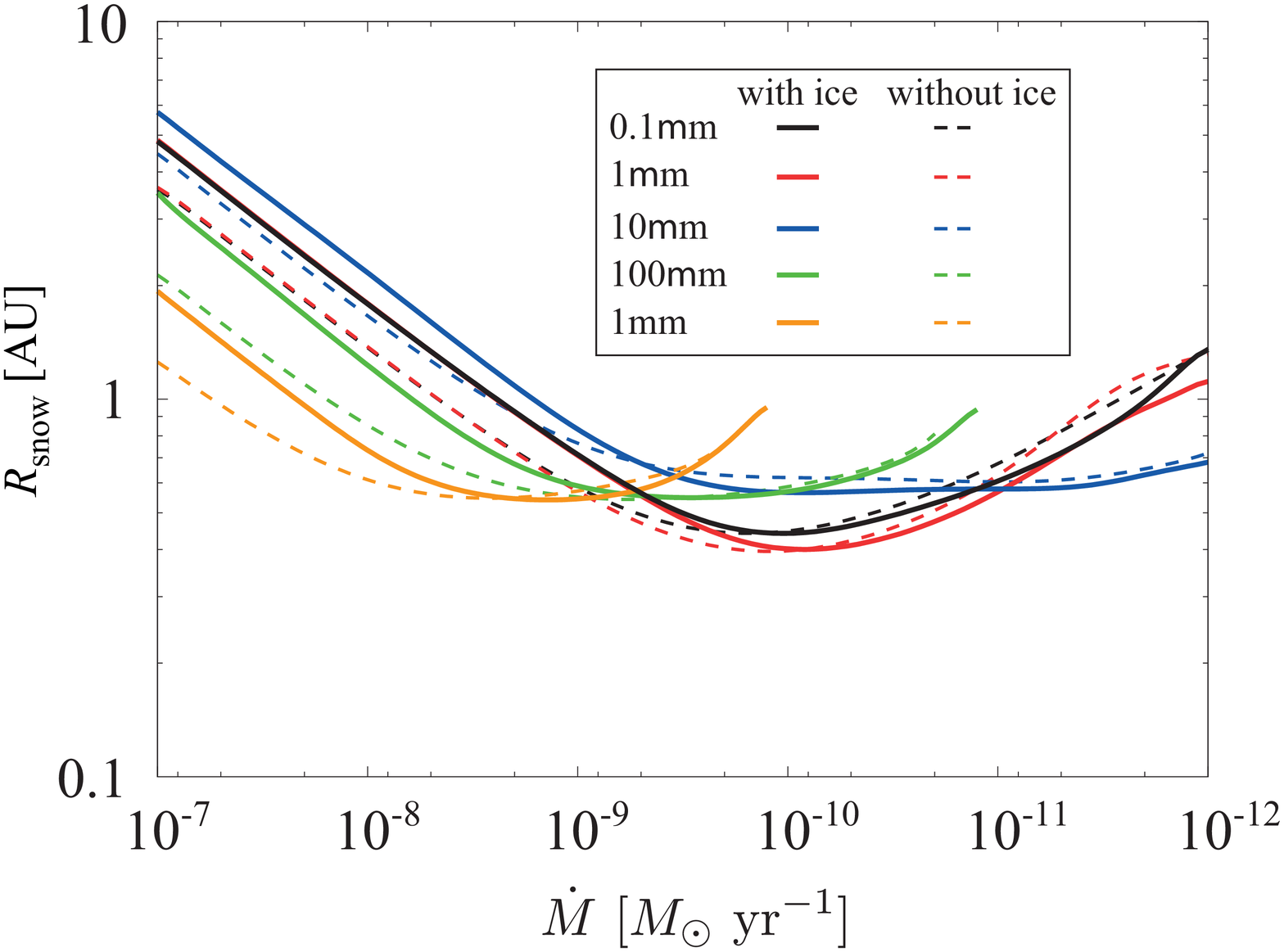}
\caption{}
\label{solar_parameter_snowline}
\end{figure}

\begin{figure}
\epsscale{1.0}
\plotone{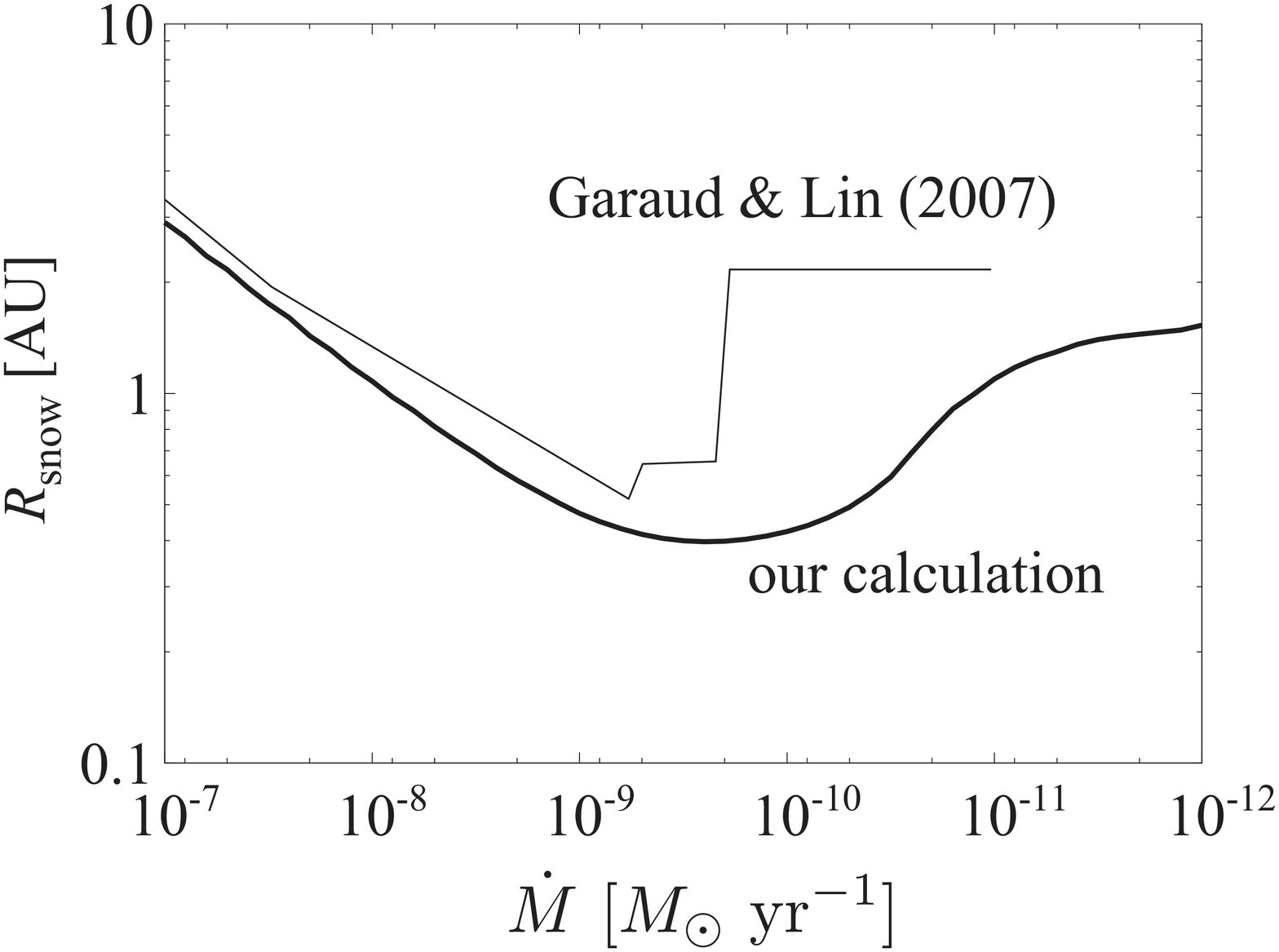}
\caption{}
\label{Garaud_Lin_snowline}
\end{figure}

\begin{figure}
\epsscale{1.0}
\plotone{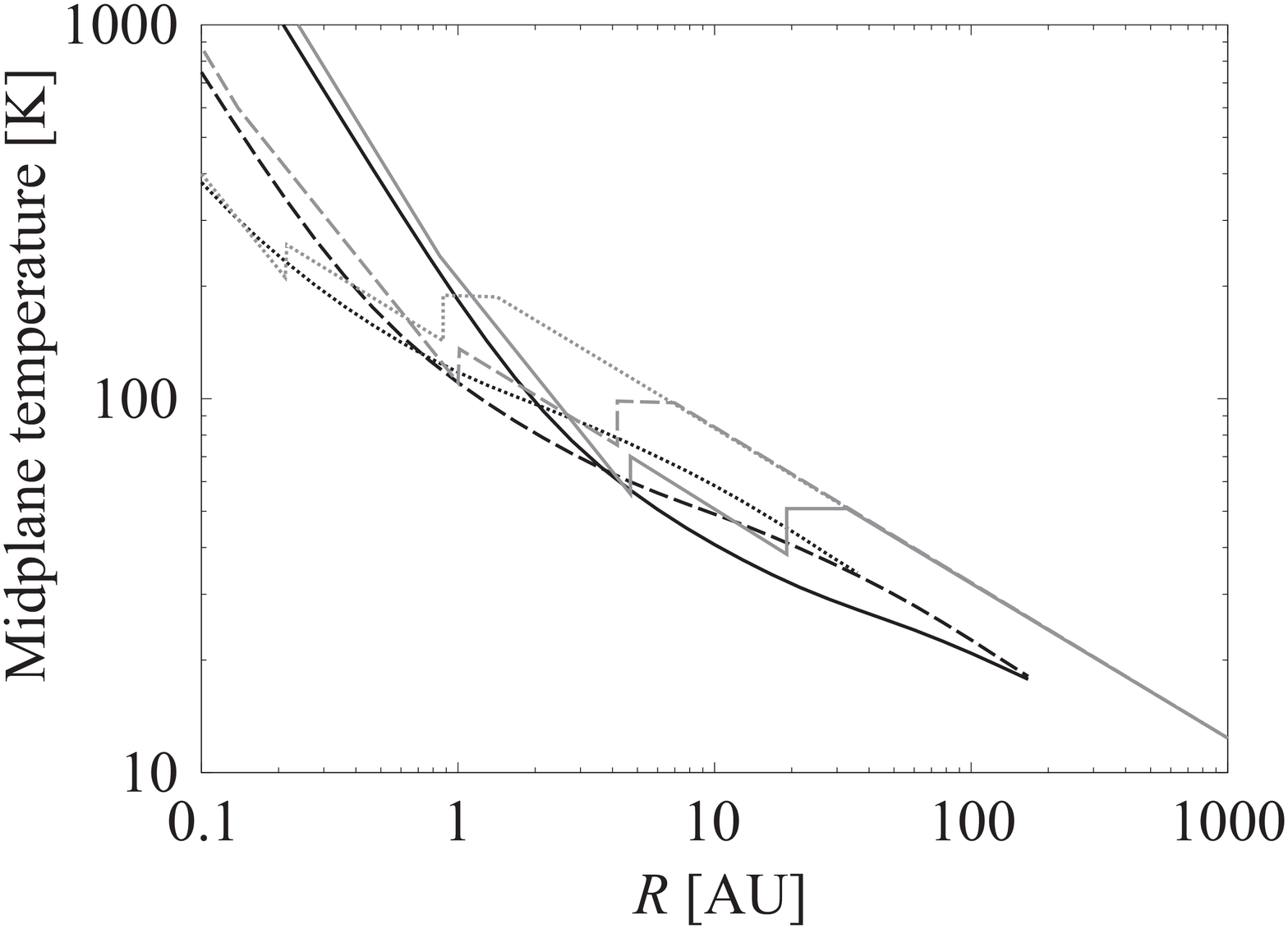}
\caption{}
\label{temperature_Garaud_Lin}
\end{figure}

\begin{figure}
\epsscale{1.0}
\plotone{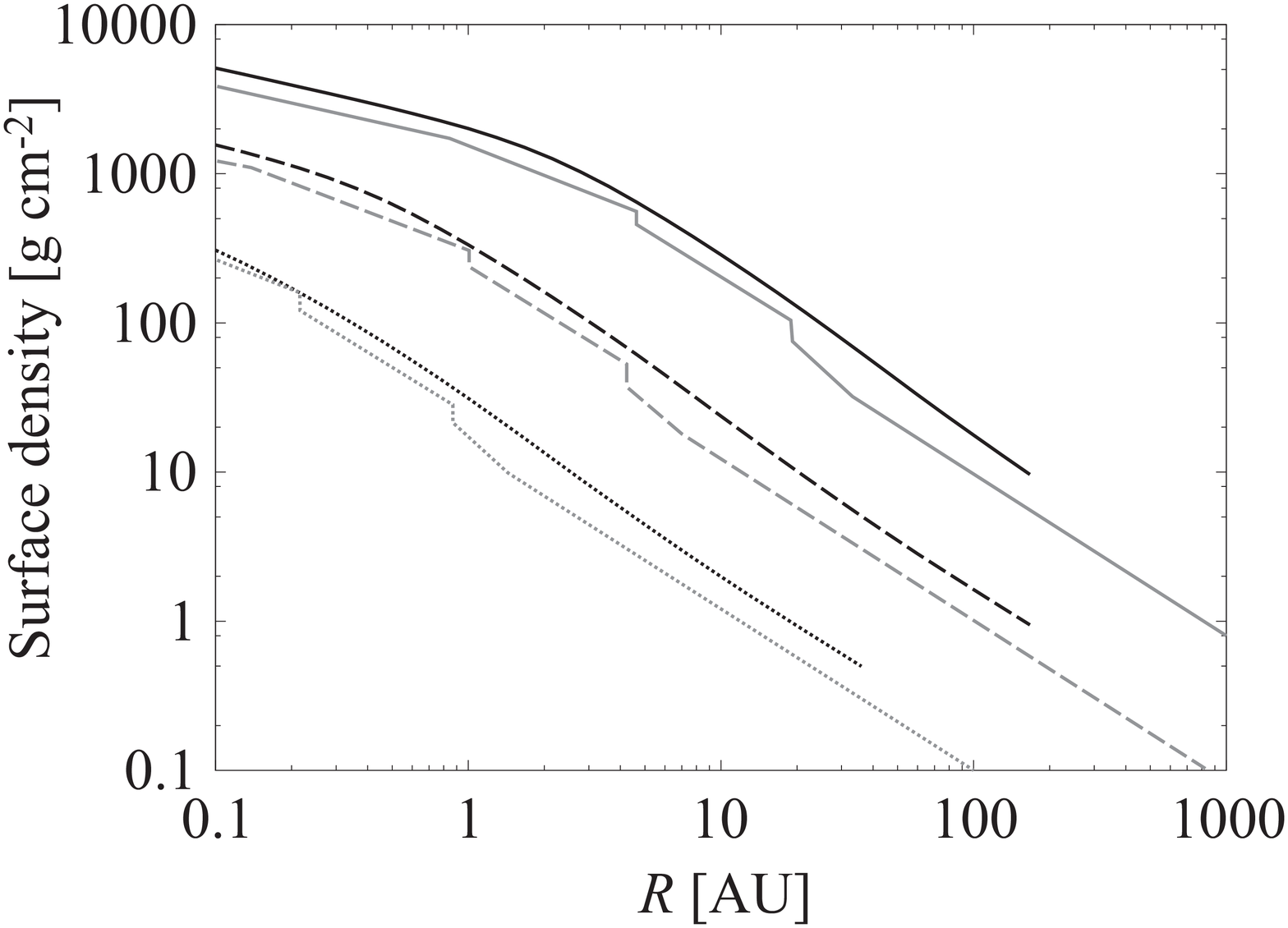}
\caption{}
\label{surface_density_Garaud_Lin}
\end{figure}

Fig. 3.--- Same as Fig. \ref{temperature_profile8_0} except the mass accretion rate and the heliocentric distances of the panels. 
The mass accretion rate is $10^{-10}M_{\odot}\mathrm{yr}^{-1}$ and the heliocentric distances of the panels (a)-(d) are 0.2, 0.4, 0.6, and 0.8 AU, respectively. 

Fig. 4.--- Ice-condensing region in the disk as a function of the mass accretion rate. 
Parameters are $M_{*} = 0.5$ $M_{\odot}$, $R_{*} = 2$ $R_{\odot}$, $T_{\mathrm{eff}} = 3000$ K,
$\alpha = 0.01$, and the dust grain size is 0.1 $\mathrm{\mu m}$. 
Panels (a)-(f) show the ice-condensing regions with $\dot{M}$ labeled in each pane. 
The horizontal and vertical axes represents the distance from the central star on the midplane and the altitude from the midplane, respectively. 
The blue and light blue regions represent the ice-condensing regions with and without the ice opacity. 
The black, red, and green curves represent the condensation front, the altitude of $H_{s}$, and the pressure scale height, respectively. 
The solid and dotted curves with each color represent the results with and without the ice opacity, respectively. 

Fig. 5.--- Heliocentric distance of the snow line as a function of the mass accretion rate. 
Parameters are $M_{*} = 0.5$ $M_{\odot}$, $R_{*} = 2$ $R_{\odot}$, $T_{\mathrm{eff}} = 3000$ K,
$\alpha = 0.01$, and the dust grain size is 0.1 $\mathrm{\mu m}$. 
The horizontal and vertical axes represent the mass accretion rate and the snow line location, respectively. 
The solid and dashed curves represent the results with and without the ice opacity, respectively.

Fig. 6.--- Heliocentric distance of the snow line as a function of the mass accretion rate with various dust grain sizes. 
Parameters are $M_{*} = 1$ $M_{\odot}$, $T_{\mathrm{eff}} = 4000$ K, $L_{*} = 1$ $L_{\odot}$, and $\alpha = 0.01$. 
The horizontal and vertical axes are the same as Figure 5. 
The black, red, blue, green, and orange curves represent the results with the dust grain size of $0.1\mathrm{\mu m}$, $1\mathrm{\mu m}$, $10\mathrm{\mu m}$, and $100\mathrm{\mu m}$, respectively. 
The solid and dashed curves represent the results with and without the ice opacity, respectively. 

Fig. 7.--- Comparison of the snow line location obtained in this study (thick curve) with that from Garaud \& Lin (2007; thin curve). 
Parameters are $M_{*} = 1$ $M_{\odot}$, $R_{*} = 1$ $R_{\odot}$, $L_{*} = 1$ $L_{\odot}$, $\alpha =  0.001$, and the dust opacity is similar to that of Garaud \& Lin (2007). 
The horizontal and vertical axes are the same as Figure 5. 

Fig. 8.--- Comparison of the midplane temperature obtained in this study (black curves) with that from Garaud \& Lin (2007; gray curves). 
Parameters are the same as Figure 7. 
The horizontal and vertical axes represent the heliocentric distance on the midplane and the midplane temperature, respectively. 
The solid, dashed, and dotted curves represent the results with the mass accretion rates of $10^{-8}M_{\odot}\mathrm{yr}^{-1}$, $10^{-9}M_{\odot}\mathrm{yr}^{-1}$, and $10^{-10}M_{\odot}\mathrm{yr}^{-1}$, respectively. 

Fig. 9.--- Comparison of the gas surface density obtained in this study (black curves) with that from Garaud \& Lin (2007; gray curves). 
Parameters are the same as Figure 7. 
The horizontal and vertical axes represent the heliocentric distance on the midplane and the gas surface density, respectively. 
The solid, dashed, and dotted curves represent the results with the mass accretion rates of $10^{-8}M_{\odot}\mathrm{yr}^{-1}$, $10^{-9}M_{\odot}\mathrm{yr}^{-1}$, and $10^{-10}M_{\odot}\mathrm{yr}^{-1}$, respectively. 

\clearpage




\clearpage

\clearpage

\end{document}